\documentclass[epj]{svjour}
\usepackage{graphics}
\usepackage{epsfig,latexsym}
\usepackage{dcolumn}% Align table columns on decimal point

\newcommand{\be}{\begin{eqnarray}}
\newcommand{\ee}{\end{eqnarray}}

\begin{document}
\title{Heavy Quark Potentials and Quarkonia Binding }
\author{P{\'e}ter Petreczky % etc
}                     % Do not remove
\institute{
Nuclear Theory Group, Department of Physics,
Brookhaven National Laboratory, Upton, New York 11973-500, USA}
\date{}
\abstract{
I review recent progress in studying in-medium 
modification of inter-quark forces at finite 
temperature in lattice QCD. Some applications to
the problem of quarkonium binding in potential 
models is also discussed.
 % end of PACS codes
} %end of abstract
\maketitle

%%%%%%%%%%%%%%%%%%%%%%
\section{Introduction}
%%%%%%%%%%%%%%%%%%%%%%%

The study of in-medium modifications of inter-quark forces at
high temperatures is important for detailed theoretical 
understanding of the properties of Quark Gluon Plasma as well to
detect its formation in heavy ion collisions. In particular, it
was suggested by Matsui and Satz that color screening at high temperature
will result in dissolution of quarkonium state and the corresponding
quarkonium suppression could be a signal of Quark Gluon Plasma 
formation \cite{MS86}. 

Usually the problem of in-medium modification of inter-quark
forces is studied in terms of so-called finite temperature
heavy quark potentials, which are, in fact, the differences in the
free energies of the system with static quark anti-quark pair and
the same system without static charges. Alternatively this problem
can be studied in terms of finite temperature 
quarkonium spectral functions \cite{umeda02,asakawa04,datta04}
which were also discussed during this conference by Karsch, Hatsuda and
Petrov \cite{karschthis}.  
Recently substantial prog\-ress has been made in studying the free
energy of static quark anti-quark pair which I am going to
review in  the present paper. An important question is what can we
 learn about the quarkonium properties from the free energy of
static charges {which will be discussed at the end of the paper.

%%%%%%%%%%%%%%%%%%%%%%%%%%%%%%%%%%%%%%%%%%%%
\section{The free energy of static charges}
%%%%%%%%%%%%%%%%%%%%%%%%%%%%%%%%%%%%%%%%%%%%

Following McLerran and Svetitsky the partition function of the 
system with static quark anti-quark ($Q \bar Q$) pair 
at finite temperature $T$ can be written
as
\begin{equation}
Z_{Q \bar Q}(r,T)= \langle W(\vec{r}) W^{\dagger}(0) \rangle Z(T),
\end{equation}
with $Z(T)$ being the partition function of the system without static
charges and 
\begin{equation}
W(\vec{x})={\cal P} \exp( i g \int_0^{1/T} d \tau A_0(\tau, \vec{x}) )
\end{equation}
is the temporal Wilson line. $L(\vec{x})={\rm Tr} W(\vec{x})$ is also
known as Polyakov loop
and in the case of pure gauge theory it 
is an order parameter of the deconfinement
transition. As the $Q \bar Q$ pair can be either in color singlet or octet
state one should separate these irreducible contributions to the 
partition 
function. This can be done using the projection operators $P_1$ and 
$P_8$ onto color singlet and octet states  introduced in Refs. 
\cite{brown79,nadkarni86}. Applying $P_1$ and $P_8$ to
 $Z_{Q \bar Q}(r,T)$ we get the following expression for the singlet
and octet free energies of the static $Q \bar Q$ pair
\begin{eqnarray}
&
\displaystyle
\exp(-F_1(r,T)/T)=\frac{1}{Z(T)}
\frac{{\rm Tr} P_1 Z_{Q \bar Q}(r,T)}{{\rm Tr P_1}} \nonumber\\
\displaystyle
&
=\frac{1}{3} {\rm Tr} \langle W(\vec{r}) W^{\dagger}(0) \rangle \\[2mm] 
&
\displaystyle
\exp(-F_8(r,T)/T)=\frac{1}{Z(T)}
\frac{{\rm Tr} P_8 Z_{Q \bar Q}(r,T)}{{\rm Tr} P_8} \nonumber \\[1mm]
&
\displaystyle
{\bf =} \frac{1}{8} \langle {\rm Tr} W(\vec{r}) {\rm Tr} W^{\dagger}(0) \rangle
-\frac{1}{24} {\rm Tr} \langle W(\vec{r}) W^{\dagger}(0) \rangle .
\end{eqnarray}
Although usually $F_{1,8}$ is referred to as the free energy of
  the
static 
$Q \bar Q$ pair, it is important to keep in mind that it refers to the 
difference between the free energy of the system with static quark
anti-quark pair and the free energy of the system without static charges.

As $W(\vec{x})$ is a not gauge invariant operator we have to fix a gauge
in order to define $F_1$ and $F_8$. As we want that $F_1$ and $F_8$ 
have a meaningful zero temperature limit we better to fix 
the Coulomb gauge because in this gauge a transfer matrix can be defined
and the free energy difference can be related to the interaction 
energy of a static $Q \bar Q$ 
pair at zero temperature ($T=0$).  
Another possibility discussed in Ref. \cite{ophil02}  is to replace
the Wilson line by a gauge invariant Wilson line using the eigenvector of
the spatial covariant Laplacian \cite{ophil02}. 
For the singlet free energy both
methods were tested and they were 
shown to give numerically indistinguishable results, which  in the zero 
temperature limit are the same as the canonical results obtained
from Wilson loops. The interpretation of the color octet free
energy at small temperatures is less obvious and will be discussed
separately. 
One can also define the color averaged free energy
\begin{eqnarray}
&
\displaystyle
\exp(-F_{av}(r,T)/T)=\frac{1}{Z(T)} 
\frac{{\rm Tr }(P_1+P_8) Z_{Q \bar Q}(r,T)}{{\rm Tr} (P_1+P_8)}=\nonumber \\[1mm]
&
{\bf =}\frac{1}{9} 
\langle {\rm Tr}  W(\vec{r}) {\rm Tr} W^{\dagger} (0) \rangle,
\end{eqnarray}
which is expressed entirely in terms of gauge invariant Polyakov
loops. 
This is the reason why 
it was extensively studied on lattice during the last two decades.
The color averaged free energy is a thermal average over the
free energies in color singlet and color octet states
\begin{eqnarray}
&
\displaystyle
\exp(-F_{av}(r,T)/T)=\nonumber\\
&
\frac{1}{9} \exp(-F_1(r,T)/T)+
\frac{8}{9} \exp(-F_8(r,T)/T).
\label{fav18}
\end{eqnarray}
Therefore it gives less direct information about medium modification
of inter-quark forces.
Given the partition function $Z_{Q \bar Q}(r,T)$ we can calculate not
only the free energy but also the 
entropy as well as 
the internal energy of the static charges
\begin{eqnarray}
&
\displaystyle
S_i(r,T)=\frac{\partial}{\partial T} \ln \biggl( 
T \frac{Z_{Q \bar Q}^i(r,T)}{Z(T)} \biggr)=
-\frac{\partial F_i(r,T)}{\partial T}, \label{si}\\
&
\displaystyle
U_i(r,T)=T^2 \frac{\partial}{\partial T} \ln \biggl( 
\frac{Z_{Q \bar Q}^i(r,T)}{Z(T)} \biggr)\nonumber\\
&
\displaystyle
~~~~~~~~~~~~=
F_i(r,T)+T S_i(r,T), \label{ui}\\
&
~~~i=1,8,av. \nonumber
\end{eqnarray}

\begin{figure}
\includegraphics[width=6cm]{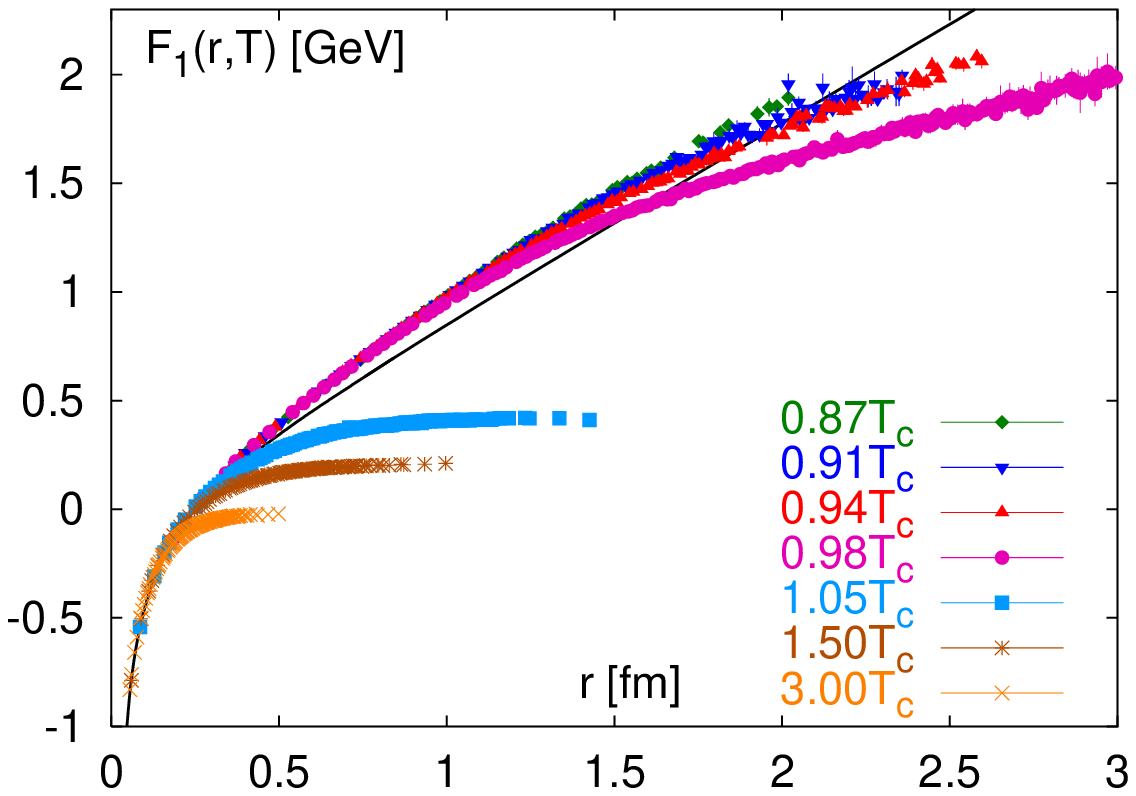}
\vskip0.3truecm
\includegraphics[width=6cm]{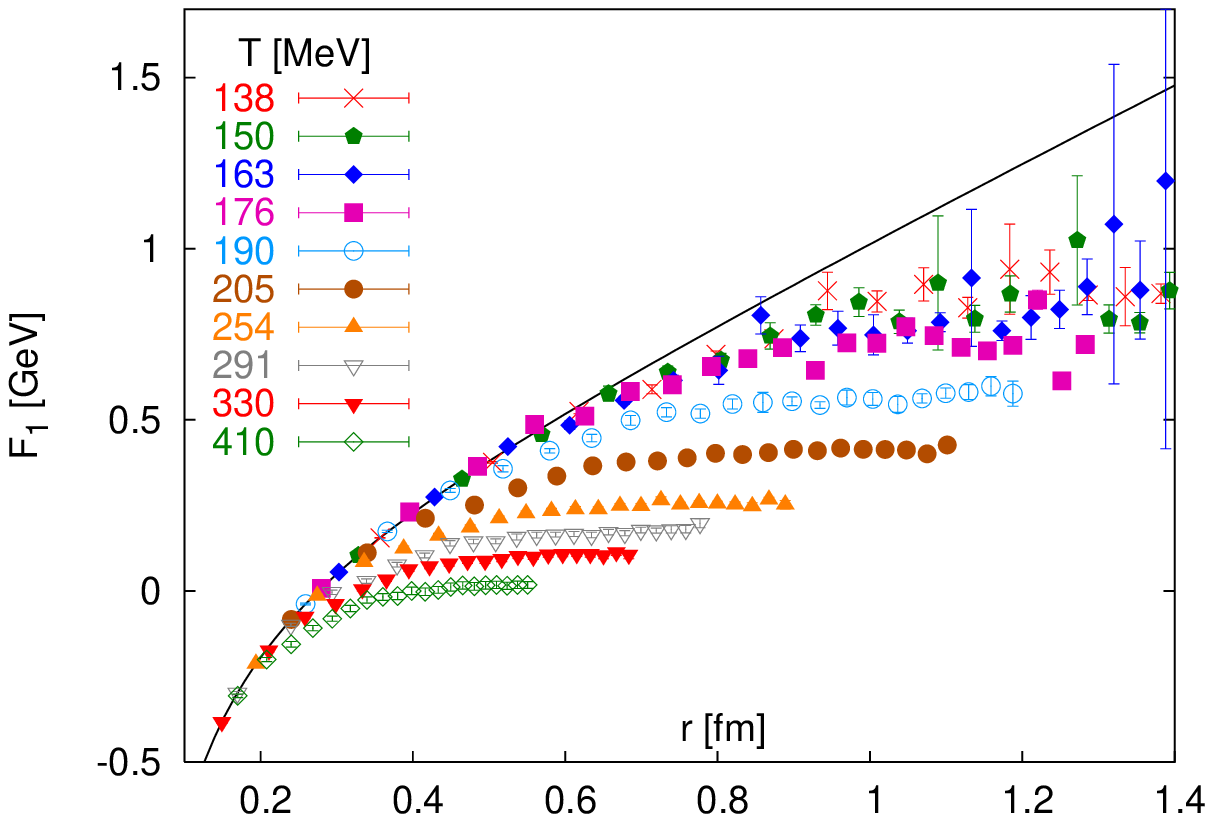}
\caption{The color singlet free energy in quenched \cite{okacz02,progres} (top)
and three flavor \cite{petrov04} (bottom) QCD. The solid black line is the $T=0$ 
singlet potential.}
\label{fig_f1}
\end{figure}
\begin{figure}
\includegraphics[width=5.9cm]{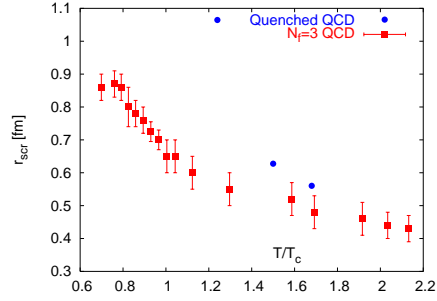}
\caption{The effective screening radius versus $T/T_c$ \cite{petrov04} .}
\label{fig_rscr}
\end{figure}

%%%%%%%%%%%%%%%%%%%%%%%%%%%%%%%%%%%%%%%%%%%%%%%%%%%%%%%%
\section{Free energy of a 
static $Q \bar Q$ pair and screening of 
inter-quark forces at high temperatures}
%%%%%%%%%%%%%%%%%%%%%%%%%%%%%%%%%%%%%%%%%%%%%%%%%%%%%%%%

Perturbatively the quark anti-quark potential can be related to the 
scattering amplitude corresponding to one gluon exchange and  in
the non-relativistic limit it is given by
\begin{eqnarray}
&
\displaystyle
V(r)=\langle T^a T^b \rangle  g^2 \int 
\frac{d^3 k}{(2 \pi)^2} e^{i \vec{k} \cdot
\vec{r}} D_{00}(k).
\end{eqnarray}
Here $D_{00}(k)$ is the temporal part of the Coulomb gauge 
gluon propagator and in general {\bf it} has the form
$$
D_{00}(k)=({\bf k}^2+\Pi_{00}({\bf k}))^{-1}.
$$
Furthermore the averaging over color gives $\langle T^a T^b \rangle=-4/3$ for
the color singlet and $\langle T^a T^b \rangle=+1/6$ for the color octet case.
At zero temperature the polarization operator $\Pi_{00}$ gives
rise only to running of the coupling constant $g=g(r)$ (recall that 
$\alpha_s=g^2/(4 \pi)$). But at finite
temperature $T$ it has a non-trivial infrared limit
$\Pi_{00}(k \rightarrow 0)=m_D^2 =g T \sqrt{N_c/3+N_f/6}$.
Therefore at distances $r \gg 1/T$ the potential has the form
\begin{equation}
V(r,T)=\langle T^a T^b \rangle \frac{g^2}{4 \pi r} \exp(-m_D r).
\end{equation} 

The singlet and octet free energies defined in the previous section 
can also be easily calculated 
in leading order perturbation theory. Again because of 
$\Pi_{00}(k=0)=m_D^2$ one has 
\begin{eqnarray}
&
\displaystyle
F_{1,8}(r,T)=(-\frac{4}{3}, \frac{1}{6}) \frac{g^2}{4 \pi r} \exp(-m_D r). 
\label{f1p}
%&
%\displaystyle
%F_8(r,T)=+\frac{1}{8} \frac{g^2}{4 \pi r} \exp(-m_D r)
%\label{f8p}
\end{eqnarray}
At leading order the singlet free energy has exactly the same form as the 
potential and has no entropy contribution. This is the reason why the free
energies of static $Q \bar Q$ pair were (mis)interpreted as potentials.
At next to leading order which is ${\cal O}(g^3)$ the free energies  have  the 
form 
\begin{equation}
F_{1,8}(r,T)=(-\frac{4}{3},\frac{1}{6}) \frac{g^2}{4 \pi r} \exp(-m_D r)
-\frac{g^2 m_D}{3 \pi},
\end{equation}
and the entropy contribution $-TS$ appears (recall Eq. \ref{si}).
For the singlet case the entropy has the form
\begin{equation}
S_1(r,T)=\frac{g^2 m_D}{3 \pi T} (1-\exp(-m_D r)).
\end{equation}
It has the asymptotic value of $\frac{g^2 m_D}{3 \pi T}$ at large 
distances and
vanishes for $rT \ll 1$. Similarly one can calculate the color
octet entropy to be 
\begin{equation}
S_8(r,T)=\frac{g^2 m_D}{3 \pi T} (1+\frac{1}{8}\exp(-m_D r)).
\end{equation}
Contrary to the color singlet case it does not vanish at small 
distances. We can also calculate the internal energy which for
color singlet state, for example, can be written as
\begin{equation}
U_1(r,T)=-\frac{4}{3} \frac{g^2}{4 \pi r} \exp(-m_D r)-
\frac{g^2 m_D}{3 \pi} \exp(-m_D r),
\end{equation}
and unlike the free energy vanishes at large distances
(at least to order $g^3$).
Using Eqs. (\ref{fav18}), (\ref{f1p}) 
one can easily get the perturbative
result for the color averaged free energy. For $rT>1$ the 
exponentials in Eq.  (\ref{fav18}) can be expanded and we arrive at
the well known leading order result \cite{mclerran81,nadkarni86}
\begin{equation}
F_{av}(r,T)=-\frac{1}{9} \frac{g^4}{(4 \pi r )^2 T} \exp(-2 m_D r).
\end{equation}
\begin{figure}
\includegraphics[width=6cm]{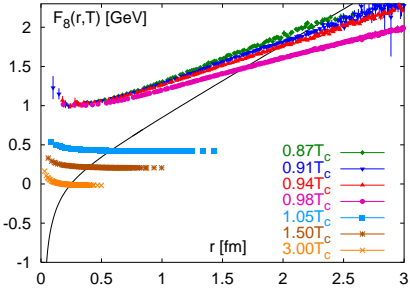}
\caption{The color octet free energy in quenched QCD \cite{progres}.
The solid black line is the zero temperature singlet potential.
}
\label{fig_f8}
\end{figure}

%%%%%%%%%%%%%%%%%%%%%%%%%%%%%%%%%%%%%%%%%%%%%%%%%
\section{Numerical results on the free energy of static $Q \bar Q$}
%%%%%%%%%%%%%%%%%%%%%%%%%%%%%%%%%%%%%%%%%%%%%%%%%

\subsection{Color singlet free energy}
In this section I am going to review recent lattice results on the
free energy of a 
static quark anti-quark pair. I will start the discussion
with the 
case of the color singlet channel. The color singlet free energy has been
extensively studied only during the last three years. Presently results
are available for SU(2) and SU(3) gauge theories 
\cite{ophil02,okacz02,digal03,okaczlat03,okacz04} as well as in two flavor 
\cite{okacz05_nf2} and 
three flavor QCD \cite{petrov04}. While for pure gauge theories these studies
are very systematic and lattice artifacts are under control, for
full QCD they are still in the exploratory stage. 

In Fig. \ref{fig_f1} the singlet free energy for SU(3) gauge theory 
(QCD without dynamical quarks or quenched QCD) is shown for different 
temperatures together
with the zero temperature quark anti-quark potential. For temperatures below
the transition temperature $T_c \simeq 270$MeV the free energy rises 
linearly with the distance $r$ signaling confinement. Above deconfinement
 $T>T_c$ the free energy has a finite value at infinite separation indicating 
screening. One can also see from the figure that at short distances, 
$rT \ll 1$,
the free energy is temperature independent in the entire temperature range and
equal to the zero temperature potential. The singlet free energy for
three flavor QCD 
is also shown in Fig. \ref{fig_f1}. The main difference compared to the case of
SU(3) gauge theory is the fact that the free energy reaches a constant value at
all temperatures. At low temperatures this is interpreted as string
breaking, the flux
tube breaks if  enough energy is accumulated to create a light
quark anti-quark pair 
which with the static $Q \bar Q$  could form a static-light meson, i.e. when 
$V(r=r_{scr})=E^{binding}_{heavy-light}$ ( see e.g. discussion in
Ref. \cite{digal01a} ). 
The distance $r_{scr}$, where the free energy effectively flattens off
depends on the 
temperature, it becomes smaller as the
temperature increases. Therefore it is
interpreted  as an  effective screening radius and is shown in Fig. 
\ref{fig_rscr}.
At low temperatures, $T<T_c$ it has a value of about $0.9$ fm and
rapidly decreases
near the transition point. While close to $T_c$ the effect of
dynamical quarks is important for
the value of the screening radius, at high temperatures the value of
the screening radius is 
similar in quenched and full QCD.

\begin{figure}
\includegraphics[width=6.0cm]{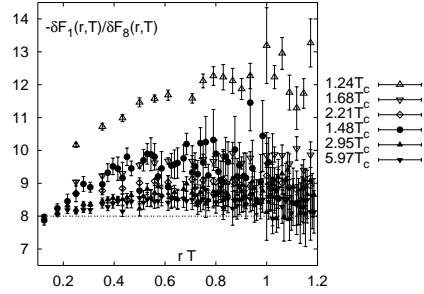}
\caption{The ratio of color singlet and color octet free energies \cite{progres}.}
\label{fig_rat18}
\end{figure}

\subsection{Color octet free energy}

The color octet free energy is shown in Fig. \ref{fig_f8} for quenched
QCD. At short distances it is repulsive as expected from
 perturbation theory. Above the deconfinement
temperature it has strong temperature dependence which presumably
comes from the
entropy contribution and is present even at short distances.
At high temperatures this is also expected from perturbation theory
(see previous section).
Above deconfinement the color octet free energy has the
same large distance asymptotic value as the color singlet free energy
$F_8(r \rightarrow \infty,T)=F_1(r \rightarrow \infty,T)=F_{\infty}(T)$.
This is intuitively expected, at large distances the quark and
anti-quark are 
screened
by their respective ``clouds'' and do not know anything about their
relative 
color orientation.
One should note that also  below $T_c$ the octet and singlet free
energies 
become
equal at large distances ( compare Figs. \ref{fig_f1} and
\ref{fig_f8}) though 
there is
no particular physical reason for this. I will discuss this problem at
the end 
of this 
subsection. These features of the color octet free energy are present
also in 
the case
of three flavor QCD \cite{petrov04}.

Perturbation theory predicts that at high temperatures
we expect for the ratio of  
$\delta F_{1,8}(r,T)=(F_{1,8}(r,T)-F_{\infty}(T))$
the 
following   $\delta F_1(r,T)/F_8(r,T) \simeq -8$.
In Fig. \ref{fig_rat18}
the lattice data in SU(3) gauge theory are confronted with these
expectations. 
As one
can see from the figure the data for this ratio are close to $-8$ for 
temperatures
$T>2T_c$. 

\begin{figure}
\includegraphics[width=6cm]{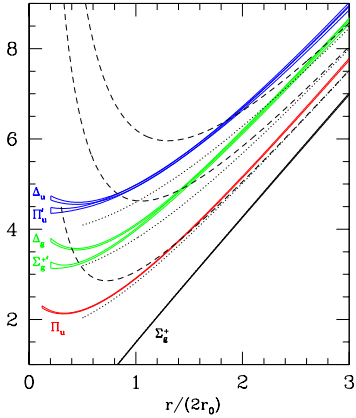}
\caption{The hybrid potentials in quenched QCD from Ref. \cite{morningstar99}.
$\Sigma_g^{+}(r)=V(r)$ is the singlet potential and the dashed lines
  are 
predictions
from string model.}
\label{fig_hyb}
\end{figure}

While at high temperatures the  meaning of the notion
 of color octet free energy is clear, the 
meaning of this quantity at low temperatures (``confinement region'') is less 
evident. To understand the problem let us first discuss the spectrum
 of static quark
anti-quark free energies at zero temperature. The lowest energy 
 level  of a
static $Q \bar Q$ pair is the one where the quark and anti-quark are
 color singlet state.
{The corresponding energy as a function of the quark anti-quark
 separation
 is the singlet static potential or simply the static potential
 determined 
in terms
of Wilson loop and used extensively in potential models (see Refs. 
\cite{bali01,yellow}).
There  are 
also higher energy levels, whose energy functions are
 called hybrid potentials for which the gluon 
fields between the static charges are in excited state (or in other words the 
string formed between the quark and anti-quark is excited) 
\cite{bali01,morningstar99,juge03}. 
The spectrum of hybrid potentials is shown in Fig. \ref{fig_hyb}. 
The hybrid potentials are labeled by the angular momentum 
 projection of the gluon field
configurations on the quark anti-quark axis, $L=0,1,2$ (denoted as 
$\Sigma$, $\Pi$, $\Delta$), CP (even, g, or odd, u) and the reflections
 properties with respect
to the
plane passing through the quark anti-quark axis (even,+, or odd, -) 
\cite{bali01}. 
The most
distinct feature of the hybrid potentials is the different slope at
 small 
distances,
i.e. in contrast to the singlet potential the hybrid potentials are repulsive.
They have a shape which is similar to the shape of the octet free energy at low
temperatures shown in Fig. \ref{fig_f8}. It has been shown that at
 short 
distances 
hybrid potentials correspond to the perturbative color octet potential 
\cite{brambilla00}. 
This means 
that at short distances hybrid potentials correspond to a state in
 which the 
static 
$Q \bar Q$ pair is in octet state and the net color charge is
 compensated by 
soft gluons
field which make the whole object color singlet (obviously only
 singlet 
objects can exist
in the confinement region). At very small distances the soft gluon
 field is 
decoupled and 
the energy is dominantly determined by the 
large repulsive interaction of the
static quark and anti-quark \cite{brambilla00}.

The correlators which enters the definition of the 
color singlet and octet free energy
have the following spectral representation \cite{jahn04}
\begin{eqnarray}
&
\langle {\rm Tr} W(\vec{r}) {\rm Tr} W^{\dagger}(0) \rangle = \sum_n e^{-E_n(r,T)/T} \\
&
\langle {\rm Tr} W(\vec{r})  W^{\dagger}(0) \rangle =\sum_n c_n e^{-E_n(r,T)/T},
\end{eqnarray}
\begin{figure}
\includegraphics[width=6cm]{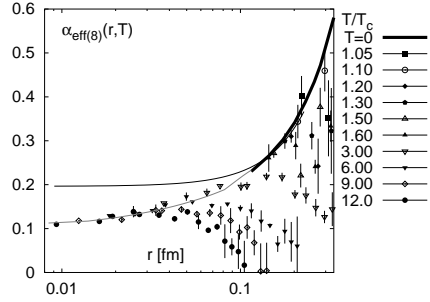}
\caption{The quenched running coupling constant at finite temperature
  in the color 
octet  case \cite{progres}. The thick black line represents the
  lattice data at zero
temperature on the running coupling. The thin black line is the
  running coupling derived
from Coulomb plus linear parametrization of the zero temperature
  potential. Finally the thin
gray line is the 3-loop running coupling in qq scheme \cite{necco01}. }
\label{fig_alpha}
\end{figure}
where $E_n$ denotes the
different energy levels of the quark anti-quark system: singlet 
potential, hybrid potentials, singlet potentials plus glueballs, etc. The weights $c_n$
in general are
different from one and may have non-trivial $r$-dependence \cite{jahn04}.
Because of asymptotic freedom $c_1$ should approach unity at short distances, while
$c_{n>1}$ should vanish; at short distances perturbation theory can be applied and the
correlator $\langle {\rm Tr} W(\vec{r})  W^{\dagger}(0) \rangle$ gives the singlet potential.
The tendency of $c_1$ approaching unity is clearly seen in the lattice data presented in
Ref. \cite{jahn04}. Since the gap between the zero temperature (ground state) potential
and the lowest hybrid potential is large for not too large distances the color octet
free energy is expected to be given by
\begin{equation}
e^{-F_8(r,T)/T} \simeq \frac{1-c_1(r)}{N^2-1} e^{-E_1(r)/T},~~E_1(r)=V(r),
\end{equation}
i.e. it is determined by the singlet potential $V(r)$. 
This is the reason why at very large distances the color singlet and color octet
free energies are equal in the low temperature region (compare Fig. \ref{fig_f1}
and Fig. \ref{fig_f8}).
On the other hand for
sufficiently small distances $1-c_1(r) \simeq 0$ and the octet free energy is 
expected to be determined by the lowest lying hybrid potential $F_8(r,T) \simeq E_2(r)+T \ln8$. 

\begin{figure}
\includegraphics[width=6cm]{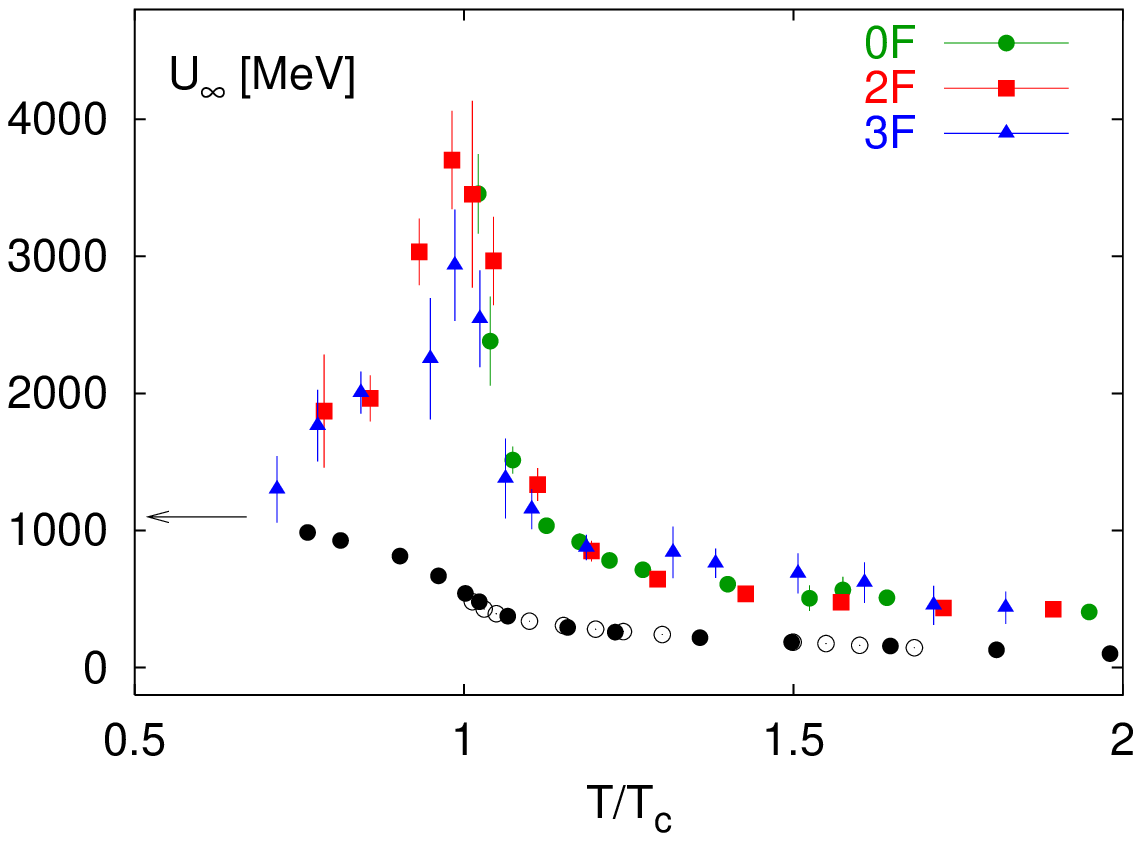}
\includegraphics[width=6cm]{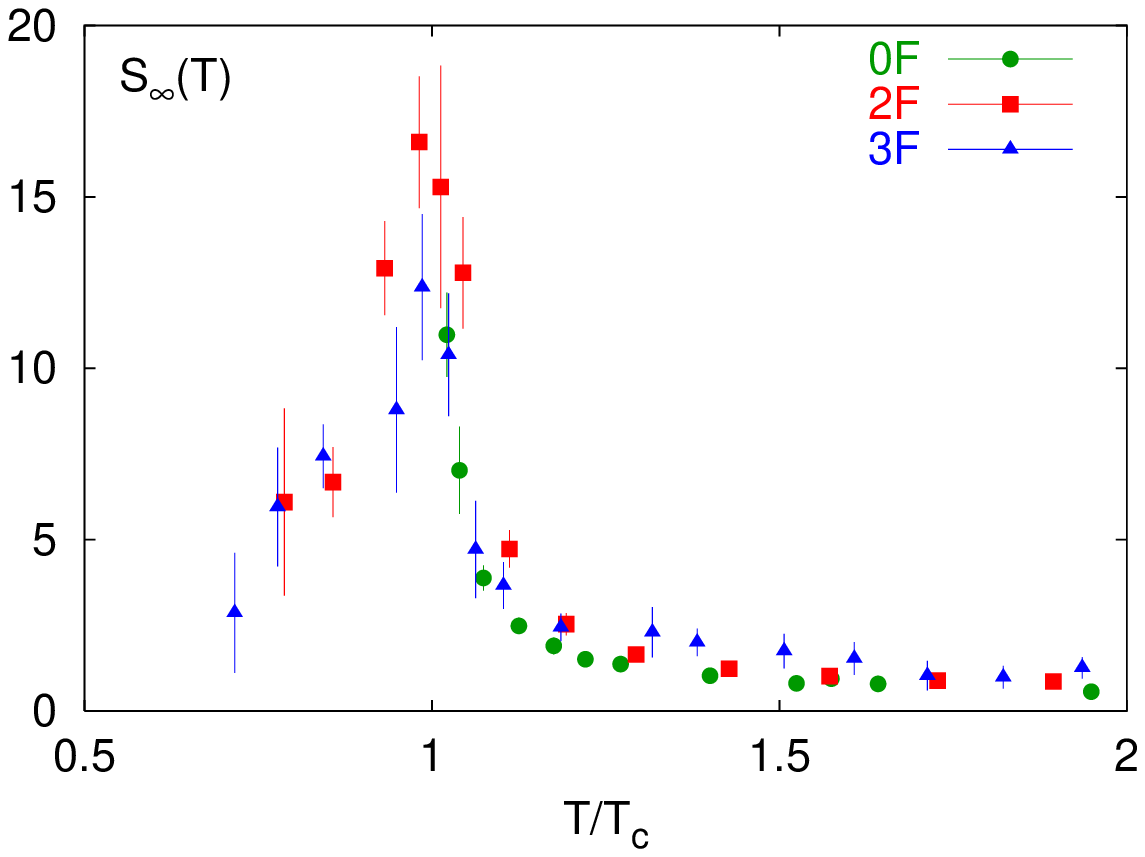}
\caption{The internal energy (to) and entropy (bottom) of static $Q
  \bar Q$ pair at infinite separation 
in quenched \cite{progres}, 2 flavor \cite{okacz05_nf2} and 3 flavor
  QCD \cite{petrov04}. The filled and open black circles is the free energy
in 2 flavor and quenched QCD respectively.}
\label{fig_suinf}
\end{figure}

\begin{figure}
\includegraphics[width=7cm]{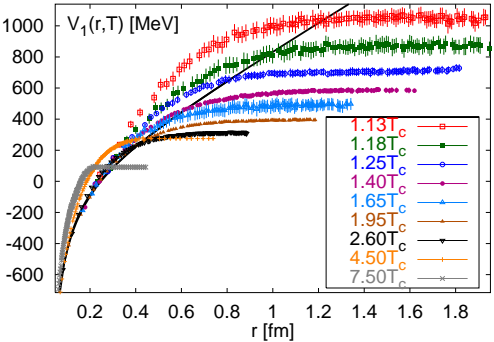}
\caption{The internal energy in quenched QCD \cite{okaczlat03}. The
  solid black line is the $T=0$ potential.}
\label{fig_u1}
\end{figure}

\subsection{Running coupling constant at finite temperature}
To study how the free energies approach the zero temperature limit as
well as to make contact 
with perturbation theory at short distances it is convenient to
introduce the effective running coupling constant
$\alpha_{eff}(r)$. This quantity can be also used to quantify the
strength of interaction at least within the
perturbative framework. At zero temperature the most convenient way to 
introduce the effective 
running coupling constant is through the force between quark and anti-quark, 
$\alpha_{eff}(r)=(3/4) \cdot r^2 (dV/dr)$.  This definition avoids
many problems in the perturbative 
calculation of the effective coupling constant \cite{necco01}.
One can define similar quantities at finite temperatures 
\cite{okacz04,zantowthis}:
$\alpha_{eff\ 1,8}(r,T)=(3/4, -6) \cdot r^2 (d F_{1,8}(r,T)/d r)$
For the color singlet case in quenched QCD 
$\alpha_{eff}$ was discussed in detail in Ref. \cite{okacz04}, where it was also 
pointed out
that $\alpha_{eff 1} \simeq \alpha_{eff 8}$. In Fig. \ref{fig_alpha}
the running coupling constant is shown for
the octet case (the results in the singlet case are essentially
identical). At short distances 
the effective coupling constant is temperature independent and
coincides 
with the zero temperature  result.
At larger distances its deviates from the zero temperature result and
after approaching a maximum it 
drops because of the onset of screening. Note that for
temperatures close to
{but already above} $T_c$ the 
effective running coupling constant follows the zero temperature
running coupling up to distances 
of about $0.3$ fm. At such distances the running coupling is not
controlled by the perturbation theory but rather by the
linear part of the potential which gives the nearly quadratic rise of
this quantity. Thus some non-perturbative 
confining physics survives deconfinement.  

\section{Entropy and internal energy of a static quark anti-quark pair}
As it was noticed in section 2, given the partition function $Z_{Q \bar Q}$ 
we can calculate the entropy and internal energy difference of the system with static charges
and the same system without them, which for the sake of simplicity is called entropy and
internal energy of  $Q \bar Q$. Using Eqs. (\ref{si}) and (\ref{ui})  the entropy and internal energy
has been calculated for infinite separation in quenched
\cite{okaczlat03}, two flavor \cite{okacz05_nf2}
and three flavor \cite{petrov04} QCD.   
The results are shown in Fig. \ref{fig_suinf}.
Both the entropy and the internal energy show a very large 
increase near $T_c$. For quenched QCD the internal energy 
has been calculated for any separation $r$ 
\cite{okaczlat03} and the results are shown in Fig. 
\ref{fig_u1}. One can see that with the exception of
the small distance region where the internal energy coincides 
with the zero temperature potential the internal energy
is larger than the free energy. In the high temperature 
limit the internal energy has no constant piece at large
distances proportional to the temperature,  therefore it is tempting 
to interpret the free energies as potentials. However, the 
large increase of $U(r,T)$ near $T_c$ make such interpretation problematic.

\section{Quarkonium binding at finite temperature}
Following the suggestion by Matsui and Satz \cite{MS86} 
the problem of quarkonium binding at finite 
temperature has been studied using potential models with 
some phenomenological 
screened potential (see e.g. \cite{karsch88,ropke88,hashimoto88} ) 
which led to the conclusions that
$J/\psi$ dissolves in the Quark Gluon Plasma at temperatures close 
to $T_c$. More recently the free
energy has been used as the potential in the Schroedinger 
equation and it was found that $J/\psi$ can
survive only to temperatures $1.1T_c$ \cite{digal01b}. 
As the free energy contains an $r$-dependent
entropy contribution the validity of this approach is doubtful.  
Finally, very recently the internal
energy calculated in Ref. \cite{okaczlat03} was used 
as a potential in Schroedinger equation
\cite{shuryak04} and it was found 
that $J/\psi$ can survive till $1.7T_c$ which is
not inconsistent with lattice calculations of $J/\psi$ 
spectral function \cite{asakawa04,datta04}.  
However, to test the validity 
of potential models it is not sufficient to make 
statements about the dissolution temperature of a given
quarkonium state, one should investigate the 
change in the properties of heavy quarkonium bound states.
For example, potential models with screening 
predict a decrease in the quarkonium masses. The most convenient
way to compare the prediction of potential 
models with direct calculation of quarkonium spectral functions
is to calculate the Euclidean meson 
correlator at finite temperature (see contribution by M{\'o}csy to 
this proceedings \cite{mocsyhard04}). This quantity 
can be reliably calculated on the lattice.
The basic idea is to use the model spectral function 
%\begin{equation}
$$
\sigma(\omega,T)=\sum_i 2 F_i (T)\delta(\omega^2-M_i^2(T))
+m_0 \theta(\omega-s_0(T)) \omega^2 \nonumber
$$
%\end{equation} 
containing bound states (resonances) and continuum 
\cite{shuryak93,mocsyhard04}. 
For a given screened potential 
one can solve the \\
Schroedinger equation 
and determine the radial wave function (or its derivative) 
at the origin $R_i(0)$ and the binding energy $E_i$.  
The parameters of the model spectral functions
can be related to these quantities, $F_i(T) 
\sim | R(0)|^2$, $M_i=2m_{c,b}+E_i$ and acquire temperature
dependence because of the temperature dependence of the potential. 
Here $m_{c,b}$ is the constituent quark
mass of c- and b-quarks. The threshold of the continuum $s_0(T)$ 
can be related to the asymptotic value of the potential
at infinite distance $s_0(T)=2 m_{c,b}+V_{\infty}(T)$. 
Having specified completely the spectral function of the model the 
Euclidean correlator can be calculated
\begin{equation}
G(\tau,T)=\int_0^{\infty} d \omega \sigma(\omega,T) 
\frac{\cosh(\omega (\tau-1/(2T)))}{\sinh(\omega/(2T))}
\end{equation} and compared
with lattice results \cite{mocsyhard04}. 
The Euclidean correlator can be reliably calculated 
on lattice while extracting the spectral function from
it is quite difficult.
Using this approach some of the features of the quarkonium correlators,
e.g. the enhancement of the scalar correlator above 
deconfinement temperature observed in lattice
calculation, can be understood. More detailed discussion 
on this topic is given in Ref. \cite{mocsyhard04}.

\section{Conclusions}
Free energies of static $Q \bar Q$ pairs have been 
extensively studied on lattice and  provide 
useful   tool to study in-medium modification of inter-quark forces. Many body effects exert large 
influence on the free energy thus making the simple 
picture, where the temperature dependence of the free
energy reflects the screening of the two-body 
potential, not applicable. Many body effects 
are most prominent close to the transition temperature. 
For temperatures not too close to the
transition temperature free energies of static quark 
anti-quark pairs could provide useful 
qualitative (though not quantitative) insights into the 
problem of quarkonium binding in Quark Gluon Plasma.
It is interesting to note in this respect that a new model
approach to the problem of heavy quark potentials at finite
temperature was recently proposed in Ref. \cite{simonov}.

\vskip-0.5truecm
%%%%%%%%%%%%%%%%%%%%%%%%%%%
\section*{Acknowledgements}
\vskip-0.2truecm
This work was partly supported by
U.S. Department of Energy under contract DE-AC02-98CH10886. 
P.P is a Goldhaber and RIKEN-BNL Fellow. The authour would like to
thank F. Zantow for correspondence and A. Patk\'os for careful reading of the 
manuscript and valuable suggestions.

\vskip-0.5truecm
%%%%%%%%%%%%%%%%%%%%%%%%%%%

\end{document}